\documentclass[12pt]{iopart}
\usepackage[all]{xy}
\usepackage{graphicx}
\usepackage{bm}% bold math
\usepackage{amssymb}
\usepackage{amsthm}
\newcommand{\be}{\begin{equation}}
\newcommand{\ee}{\end{equation}}
\newcommand{\bea}{\begin{eqnarray}}
\newcommand{\nn}{\nonumber}
\newcommand{\eea}{\end{eqnarray}}

\begin{document}
\title{$f(R)$ gravity and scalar-tensor theory}
\author{Thomas P.~Sotiriou\footnote[1]{sotiriou@sissa.it}}
\address{SISSA-ISAS, via Beirut 2-4, 34014, Trieste, Italy and INFN, Sezione di Trieste}

\begin{abstract}
In the present paper we will investigate the relation between scalar-tensor theory 
and $f(R)$ theories of gravity. Such studies have 
been performed in the past for the metric formalism of $f(R)$ gravity; 
 here we will consider mainly the Palatini formalism, where
the metric and the connections are treated as independent quantities. 
We will try to  investigate under which circumstances $f(R)$ theories  of gravity are equivalent to scalar-tensor theory
and examine the implications of this equivalence, when it exists.
\end{abstract}

\maketitle
\section{Introduction}
\label{intro}
Soon after General Relativity (GR) was first introduced, several attempts 
to form alternative theories  of gravity where made. Some of the most interesting 
were those based on an effort to build a more general theory 
by dropping one or more of the several assumptions of GR, like the fact that the only degrees
of freedom of the gravitational field are those of the metric, or the simplicity choice that the 
gravitational Lagrangian should be a linear function of the scalar curvature.

One of the most studied alternative theories is scalar-tensor theory, where the gravitational action 
contains, apart from the metric, a scalar field which describes part of the gravitational field \cite{brans,fuji}. 
Note, however, that this theory is still a metric theory in the following sense: the scalar field is not coupled 
directly to the matter and so matter responds only to the metric. The role of scalar field is just to intervene in the 
generation of the spacetime curvature associated with the metric \cite{cwill}. 

Another type of generalization of GR that attracted early attention was based on the modification of 
the gravitational action, by allowing it to depend in a more loose way on the curvature invariants. 
A restricted class
of such theories, where the Lagrangian is a general function of the scalar curvature, 
goes under the name $f(R)$ gravity \cite{buh}. The gravitational field is represented by the metric, as in GR.
Starting from the same action, one can also 
construct a different theory if the variational principle used is changed and  
the metric and the connections are considered as independent quantities. 
This implies of course that the connections need
no longer be the Levi-Civita connections of the metric, and the resulting theory will not 
necessarily be a metric one. This is the so-called Palatini formalism of $f(R)$ 
gravity \cite{ffr,pap,kun,fer,sotlib}. A more suitable name for this theories would be 
metric-affine theories of gravity, but this name should be used only if the theory is truly not metric,
which as we will see is not the general case.

The initial motivation for the study of all the above theories was probably theoretical completeness
and consequently the interest was rather limited. However, nowdays, physics faces new challenges in cosmology, 
the most important of which is the accelerated expansion of the universe. To explain this phenomenon within
the standard picture of General Relativity, one has to resort to the introduction of an unobserved and exotic
energy component, dark energy, which has to provide about 70\% of the energy budget of the
universe \cite{Car1,Car2}. The above theories seem to offer possible ways out of this problem 
\cite{faraoni,cap1,cap2,carroll0,noji,odirev,vollick,gianl2,sot2,ama,sot3}, and due to this they 
have received considerable recent attention. 

There also seems to be a motivation to modify the gravitational action coming from high-energy physics. The effective
classical actions of gravity, predicted when one takes into consideration quantum corrections or more fundamental
theories like String theory or M theory, seem likely to include higher order curvature correction 
and/or non-minimally coupled scalar fields (see for example \cite{uti,Buch,birr}).
Of course one has to mention here, that neither the action of $f(R)$ gravity nor that of 
scalar-tensor theories has a direct relation with these effective actions. 
It merely seems interesting 
that they contain similar terms, and consequently that they can be 
used for a qualitative study of the phenomenology of those other theories.

The relation between scalar-tensor theory and $f(R)$ gravity, and their possible equivalence, 
has been studied by many authors. The largest part of the relevant literature is devoted to 
the study of the equivalence between purely metric $f(R)$ gravity and a specific type of 
scalar-tensor theory, Brans--Dicke theory \cite{tey,wands,whitt,bar,maeda,sok,cno,ame}. Less
attention has been payed to $f(R)$ gravity in the Palatini formalism \cite{flan,olkomp,olmo,gianl3}. It should also be noted that in those studies the matter action is assumed to be independent of the connection. In the present paper we are going to
focus on $f(R)$ theories of gravity within the framework of Palatini variation (metric-affine formalism), studying their possible relation or equivalence with either 
scalar-tensor theories or purely metric $f(R)$ gravity. We will also investigate the implications of a possible dependence of the matter action on the connections, which will turn out to be crucial for equivalence between different theories. Since different aspects are well understood in each of the theories mentioned aboved, 
such a study will allow us to spread our knowledge to all three theories and to better understand some subtle issues. 

\section{Actions and field equations}
\label{form}

Let us recall the actions and the resulting field equations of the three theories which we are going to refer to here.
The action of scalar-tensor theory is
\be
\label{staction}
S_{ST}=\int d^4 x\sqrt{-g}\left[\frac{y(\phi)}{2} R-\frac{\omega(\phi)}{2} (\partial_\mu \phi \partial^\mu \phi)-U(\phi)\right]+ S_M(g_{\mu\nu},\psi),
\ee
where $R$ is the Ricci scalar of the metric $g_{\mu\nu}$ and $U(\phi)$ is the potential of the scalar field $\phi$; $y(\phi)$ and $\omega(\phi)$
 are some functions of $\phi$. $S_M$ represents the matter action, which is a function of the metric and of the matter fields $\psi$. As mentioned before,
 the matter action does not depend on the scalar field and therefore the theory described by this action is a metric theory. 
By setting $y(\phi)=\phi/\kappa$, $\omega(\phi)=\omega_0/(\kappa\phi)$ and $U=V/\kappa$, where $\kappa=8 \pi G$, we derive the action
\be
\label{bdaction}
S_{BD}=\frac{1}{2\kappa}\int d^4 x\sqrt{-g}\left[\phi R-\frac{\omega_0}{\phi} (\partial_\mu \phi \partial^\mu \phi)-V(\phi)\right]+ S_M(g_{\mu\nu},\psi).
\ee
This is the action of Brans--Dicke theory, which is obviously a sub-case of scalar-tensor theories. The free parameter
$\omega_0$ is often called the Brans--Dicke parameter. Brans--Dicke is actually
the theory that will concern us here. The field equations that one derives from action (\ref{bdaction}) by 
varying with respect to the metric and the scalar field are
\bea
\label{bdf1}
G_{\mu\nu}=\frac{\kappa}{\phi} T_{\mu\nu}+\frac{\omega_0}{\phi^2}\Big(\nabla_\mu \phi \nabla_\nu \phi 
&-&\frac{1}{2}g_{\mu\nu} \nabla^\lambda \phi \nabla_\lambda \phi \Big)+\nn\\
&+&
\frac{1}{\phi}(\nabla_\mu\nabla_\nu \phi-g_{\mu\nu} \Box \phi)-\frac{V}{2\phi}g_{\mu\nu},
\eea
\be
\label{bdf2'}
\frac{2\omega_0}{\phi}\Box\phi+R-\frac{\omega_0}{\phi^2}\nabla^\mu\phi\nabla_\mu\phi-V'=0,
\ee
where $G_{\mu\nu}=R_{\mu\nu}-\frac{1}{2}R g_{\mu\nu}$ is the Einstein tensor, 
$T_{\mu\nu}\equiv\frac{-2}{\sqrt{-g}}\frac{\delta S_M}{\delta g^{\mu\nu}}$ is 
the stress-energy tensor, $\nabla$ denotes covariant differentiation, 
$\Box\equiv \nabla^\mu\nabla_\mu$ and $A'(x)\equiv \partial A(x)/\partial x$.
One can take the trace of eq.~(\ref{bdf1}) and use the result to replace $R$ in eq.~(\ref{bdf2'}) to derive
\be
\label{bdf2}
(2\omega_0+3) \Box \phi= \kappa T+\phi V'-2V,
\ee
where $T\equiv g^{\mu\nu}T_{\mu\nu}$ is the the trace of the stress energy tensor.
Notice that eq.~(\ref{bdf2'}) implies a coupling between the scalar field and the metric, but no coupling with matter, as expected, so we should not be misled by
the presence of matter in eq.~(\ref{bdf2}): the field $\phi$ acts back on matter only through the geometry.

The action of metric $f(R)$ theories of gravity is
\be
\label{metaction}
S_{met}=\frac{1}{2\kappa}\int d^4 x \sqrt{-g} f(R) +S_M(g_{\mu\nu},\psi).
\ee
Variation with respect to the metric gives \cite{buh}
\be
\label{metf}
f'(R)R_{\mu\nu}-\frac{1}{2}f(R)g_{\mu\nu}-\nabla_\mu\nabla_\nu f'(R)+g_{\mu\nu}\Box f'=\kappa T_{\mu\nu},
\ee
which are fourth order partial differential equations for the metric.

In the metric-affine formalism, the metric, $g_{\mu\nu}$, and the connections, $\Gamma^{\lambda}_{\phantom{a}\mu\nu}$,  are independent quantities 
and  both are considered as fundamental fields.
The Ricci tensor, $\tilde{R}_{\mu\nu}$, is constructed using only the connections (the ``$\tilde{\phantom{a}}$'' is placed to
denote the difference from the Ricci tensor that one could construct using $g_{\mu\nu}$, which is of course not relevant in this theory \cite{sotlib}). The scalar curvature
$\tilde{R}$ is defined to be the contraction of
$\tilde{R}_{\mu\nu}$ with the metric, i.e. $\tilde{R}\equiv g^{\mu\nu}\tilde{R}_{\mu\nu}$.
The action of metric-affine $f(R)$ theories of gravity is
\be
\label{palaction}
S_{pal}=\frac{1}{2\kappa}\int d^4 x \sqrt{-g} f(\tilde{R}) +S_M(g_{\mu\nu}, \Gamma^{\lambda}_{\phantom{a}\mu\nu}, \psi).
\ee
Notice that, in general, the matter action is allowed to depend on the connection as well. This is the physical choice 
if one considers $\Gamma^{\lambda}_{\phantom{a}\mu\nu}$ as the true connections of spacetime, since this connection will
be associated with parallel transport and consequently will define the covariant derivative \cite{sotlib}. Additionally, since
the connection is independent of the metric, it does not have to be symmetric. It can be assumed to be symmetric, if we want to consider a theory without torsion, or it can be left non-symmetric, if we want to allow torsion \footnote{This choice obviously affects the treatment of Dirac fields whose energy tensor is not generally symmetric \cite{sotlib}.}. For what follows, we will consider a symmetric connection (torsion-less theory) but, as
we will argue later on, the result of this study will not be affect by this choice. Let us note that, for the rest of this paper, we will denote the Levi--Civita connection of the metric (Christoffel symbol) as $\left\{^\lambda_{\phantom{a}\mu\nu}\right\}$, in order to distinguish it from $\Gamma^{\lambda}_{\phantom{a}\mu\nu}$.

 The field equations are derived varying the action separately with respect to
the metric and the connections and have the form \cite{sotlib}:
\bea
\label{field1sym}
& &f'(\tilde{R}) \tilde{R}_{(\mu\nu)}-\frac{1}{2}f(\tilde{R})g_{\mu\nu}=\kappa T_{\mu\nu},\\
\label{field2sym}
& &\frac{1}{\sqrt{-g}}\bigg[-\tilde{\nabla}_\lambda\left(\sqrt{-g}f'(\tilde{R})g^{\mu\nu}\right)+\tilde{\nabla}_\sigma\left(\sqrt{-g}f'(\tilde{R})g^{\sigma(\mu}\right)\delta^{\nu)}_\lambda\bigg]=\kappa\Delta_{\lambda}^{\phantom{a}\mu\nu},
\eea
where the parentheses are used to denote symmetrization with respect to the corresponding indices and the ``$\tilde{\phantom{a}}$'' over the covariant derivatives acts as a reminder of the fact that, in this case, they are constructed using the independent connection. 
$\Delta_{\lambda}^{\phantom{a}\mu\nu}\equiv\frac{-2}{\sqrt{-g}}\frac{\delta S_M}{\delta \Gamma^{\lambda}_{\phantom{a}\mu\nu}}$ is symmetric in the indices $\mu$ and $\nu$ for
a symmetric connection. If one wants to force the matter action to be independent of the connection, this quantity should be set to zero. In this case the theory reduces to what is usually called an $f(R)$ theory of gravity in the Palatini formalism. The distinction between this theory and metric-affine gravity will become very clear when we examine the possible equivalence with scalar-tensor theory in the next section.

\section{Possible equivalence between theories}
\label{main}

Consider the action (\ref{metaction}) of metric $f(R)$ gravity. One can introduce a new field $\chi$ and write
a dynamically equivalent action \cite{tey}:
\be
\label{metactionH}
S_{met}=\frac{1}{2\kappa}\int d^4 x \sqrt{-g} \left(f(\chi)+f'(\chi)(R-\chi)\right) +S_M(g_{\mu\nu},\psi).
\ee
Variation with respect to $\chi$ leads to the equation $\chi=R$ if $f''(\chi)\neq 0$, which reproduces action (\ref{metaction}).
Redefining the field $\chi$ by $\Phi=f'(\chi)$ and setting
\be
\label{defV}
V(\Phi)=\chi(\Phi)\Phi-f(\chi(\Phi)),
\ee
 the action takes the form
\be
\label{metactionH2}
S_{met}=\frac{1}{2\kappa}\int d^4 x \sqrt{-g} \left(\Phi R-V(\Phi)\right) +S_M(g_{\mu\nu},\psi).
\ee
Comparison with the action (\ref{bdaction}) reveals that this is the action of a Brans--Dicke theory
with $\omega_0=0$. So, metric $f(R)$ theories, as has been observed long ago, are fully equivalent
with a class of Brans--Dicke theories with vanishing kinetic term \cite{tey,wands}. 

It is also known that, if one defines the conformal metric $g^E_{\mu\nu}=\Phi g_{\mu\nu}$ and performs a conformal
transformation along $\Phi=R$, together with the scalar field redefinition $\Phi=\exp(\sqrt{3\,\kappa/2}\varphi)$, one can arrive to an action describing Einstein gravity, minimally coupled to a scalar field \cite{whitt,faraoni}:
\be
\label{einframe}
S_{met}=\int d^4 x \sqrt{-g^E} \left( \frac{R^E}{2\kappa}-\frac{1}{2}(\nabla \varphi)^2-V(\varphi)\right) +S_M(g^E_{\mu\nu}/\Phi(\varphi),\psi),
\ee
where $R^E$ is the Ricci scalar of the metric $g^E_{\mu\nu}$ and $V(\varphi)$ is the potential expressed in terms of $\varphi$.
The initial action (\ref{metaction}) is said to be written in the Jordan frame, where ``frame'' should be understood 
as a set of physical variables. Actions (\ref{metactionH}) and (\ref{metactionH2}) are  written in the Jordan--Helmholtz (J--H) frame. 
After the conformal transformation, the resulting action (\ref{einframe}) is expressed in what is called the Einstein frame. 
There has been a lot of discussion about which of these frames should be identified as the physical one, i.e.
the variables of which frame are the ones that a potential experiment would measure. No such distinction can be made between
the Jordan and the J--H frame since any experiment is based on the interaction with matter. In both
frames the matter is coupled only to the metric, which coincides for both frames. 
The scalar degree of freedom of gravity, which is not coupled to matter,
is just represented in the J--H frame by a scalar field whereas in the Jordan frame it is intrinsically expressed by the functional form of $f$ \footnote{In Brans--Dicke theory one starts with action (\ref{metactionH2}) and the corresponding frame is called the Jordan frame. We are here identifying it as the J--H frame to make the distinction from action (\ref{metaction}).}.

The same is not true for the Einstein frame. The matter is minimally coupled to 
$g_{\mu\nu}$, so it will be non-minimally coupled
to $g^E_{\mu\nu}$ after the conformal transformation, since the matter action is not generically conformally invariant. Therefore, even though 
the Jordan (or the J--H) frame and the Einstein frame are mathematically equivalent, they are not physically equivalent. 
Any frame can be used to perform mathematical manipulations but one will have to return to the frame initially chosen as the 
physical one in order to interpret the results physically. There have been attempts to judge which frame should be identified
as the physical one based on theoretical arguments (see \cite{sok} and references therein). However, such arguments
do not seem satisfactory or conclusive, and probably only experiments can provide the answer \footnote[3]{In \cite{sok} it is claimed that the Einstein frame is the physical one because the scalar field in the J--H frame may violate the Dominant Energy Condition (DEC), which can also lead to problems if one attempts to define the ADM (Arnowitt--Deser--Misner) energy and use it to study the stability of the ground state. However, such considerations are burdened with subtleties as well as restrictive assumptions. For example the use of ADM energy requires asymptotic flatness, which is not true for any $f(R)$ theory, especially those used in cosmological models. Moreover, the scalar field in the J--H frame can be considered as a gravitational degree of freedom and not as a matter field. Therefore it is questionable whether a violation of the DEC, together with its consequences in ADM energy considerations, constitute a problem in this case. 
Finally, it should be noted that, as discussed in \cite{olkomp}, the arguments presented in \cite{sok} for the metric formalism do not necessarily apply to the Palatini formalism as well.} (see, for example, \cite{cno} for a proposal on how experimental data can be used for this purpose). Notice also that one can argue
that if the Einstein frame is the physical one, then matter should be minimally coupled to the metric of this frame \cite{sok}. 
If this prescription is followed, one has to first perform the conformal transformation of
 the gravitational action considered, and then
minimally couple the matter. The resulting theory will be no different from General Relativity, 
since in this case the scalar
field is demoted to a common matter field, and no room is left for modifying gravity. 
Therefore, if one wants to consider 
modifications of gravity like scalar--tensor theory or metric $f(R)$ gravity, 
either the Jordan (or equivalently the J--H) frame should be assumed to be the physical one, 
or non-minimal coupling between gravity and matter should be allowed, in order for the Einstein frame to be the physical one. We will follow 
the first path here, even though most of our results will not depend on this choice due to the mathematical equivalence of the two frames, as will become obvious latter.

Consider now the action (\ref{palaction}) for metric-affine $f(R)$ gravity and introduce the scalar field $\chi$ as before. Following the same prescription
and redefining $\chi$ by using $\Phi$ the action takes the form:
\be
\label{palactionH2}
S_{pal}=\frac{1}{2\kappa}\int d^4 x \sqrt{-g} \left(\Phi \tilde{R}-V(\Phi)\right) +S_M(g_{\mu\nu}, \Gamma^{\lambda}_{\phantom{a}\mu\nu}, \psi).
\ee
Even though the gravitational part of this action is formally the same as that of action (\ref{metactionH2}), this action
is not a Brans--Dicke action with $\omega_0=0$ for two reasons: Firstly, the matter action depends on the connection unlike Brans--Dicke theory and
secondly $\tilde{R}$ is not the Ricci scalar of the metric $g_{\mu\nu}$. Therefore, there is no equivalence
between Brans--Dicke theory and the general case of $f(R)$ theories in which the connections are independent of the metric. The reason is that the theory described by the
action (\ref{palaction}) is not a metric theory. The matter action is coupled to the connection as well, which in this case is an 
independent field. This makes the theory a metric-affine theory of gravity. 

Let us examine what will happen if we force the matter action to be independent of the connection, as is isually done in the literature \cite{flan,olmo}. The field equations can then
be derived from by eqs.~(\ref{field1sym}) and (\ref{field2sym}) as mentioned before by setting $\Delta_{\lambda}^{\phantom{a}\mu\nu}=0$:
\bea
\label{field1d0}
& &f'(\tilde{R}) \tilde{R}_{(\mu\nu)}-\frac{1}{2}f(\tilde{R})g_{\mu\nu}=\kappa T_{\mu\nu},\\
\label{field2d0}
& &\frac{1}{\sqrt{-g}}\bigg[-\tilde{\nabla}_\lambda\left(\sqrt{-g}f'(\tilde{R})g^{\mu\nu}\right)+\tilde{\nabla}_\sigma\left(\sqrt{-g}f'(\tilde{R})g^{\sigma(\mu}\right)\delta^{\nu)}_\lambda\bigg]=0,
\eea
Taking the trace of eq.~(\ref{field2d0}) and replacing the result back into the same equation, we get
\be
\label{field2d02}
\tilde{\nabla}_\lambda\left(\sqrt{-g}f'(\tilde{R})g^{\mu\nu}\right)=0.
\ee
This equation implies that the connections are the Levi-Civita connections of the metric $h_{\mu\nu}=f'(\tilde{R})g_{\mu\nu}$ \cite{fer,sotlib}.
Using the dynamical equivalence of the actions (\ref{palaction}) and (\ref{palactionH2}), we can express this result using the physical variables of the J--H frame
i.e. write $h_{\mu\nu}=\Phi g_{\mu\nu}$ Then we can express $\tilde{R}$  in terms of $R$ and $\Phi$:
\be
\tilde{R}=R+\frac{3}{2\Phi^2}\nabla_\mu \Phi \nabla^\mu \Phi-\frac{3}{\Phi}\Box \Phi.
\ee
Replacing in the action (\ref{palactionH2}) the latter takes the form:
\be
\label{palactionH2d0}
S_{pal}=\frac{1}{2\kappa}\int d^4 x \sqrt{-g} \left(\Phi R+\frac{3}{2\Phi}\partial_\mu \Phi \partial^\mu \Phi-V(\Phi)\right) +S_M(g_{\mu\nu}, \psi),
\ee
where we have neglected a total divergence. The matter action has now no dependence on $\Gamma^{\lambda}_{\phantom{a}\mu\nu}$ since this was 
our initial demand. Therefore, this is indeed the action of a Brans--Dicke theory with Brans--Dicke parameter
$\omega_0=-3/2$. Notice that for this value of $\omega_0$ eq.~(\ref{bdf2}) reduces to 
\be
\label{bdf3}
\kappa T+\Phi V'(\Phi)-2V(\Phi)=0,
\ee
which is an algebraic equation between $\Phi$ and $T$ for a given potential. The analogue of this equation in the Jordan frame 
can be found if one takes the trace of eq.~(\ref{field1d0}). One can easily see that in vacuum, where $T=0$, $\Phi$ will
have to be a constant, so the theory reduces to Einstein gravity with a cosmological constant determined by the value of $\Phi$ (see \cite{fer} for a
relevant discussion in the Jordan frame).

In order to obtain the equivalence between Brans--Dicke theory with $\omega_0=-3/2$ and metric-affine $f(R)$ gravity we had to force the matter action to be independent of the connections, i.e. to reduce the theory to what is known as $f(R)$ theory of gravity in the Palatini formalism. This led to the fact that the connections became the Levi-Civita
connections of the a metric $h_{\mu\nu}=\Phi g_{\mu\nu}$, which allowed as to eliminate the dependence
of the action on the connections. We can construct a theory where
the matter action would be allowed to depend on the connections, but the connections would be assumed
to be the Levi-Civita connections of a metric conformal to $g_{\mu\nu}$ {\it a priori}. One may be misled into thinking that such a theory could be cast into the form of a Brans-Dicke theory, since
in this case, the dependence of the action on the connections can indeed be eliminated. No mathematical computations
are required to show that this is not true. The gravitational part of the action would, of course, turn out to be the same as that of
(\ref{palactionH2d0}), if $\Phi$ (its square root to be precise) is used to represent the conformal factor. Notice, however, that since the matter action in the Jordan frame had a dependence on
the connection, in the J--H frame it would have a dependence, not only on the metric, but also on $\Phi$, because the connection
will be function of both the metric and $\Phi$. Therefore, in the J--H frame the scalar field would be coupled to matter directly,
unlike Brans--Dicke theory or scalar--tensor theory in general.

The above discussion demonstrates that it is the coupling of the connections to matter that really prevents
the action (\ref{palaction}) from being dynamically equivalent to (\ref{bdaction}). One cannot achieve such
equivalence by constraining the connection. The only exception if the conformal factor relating $g_{\mu\nu}$ and the metric that is compatible to the connection, is a constant. 
In this case the theory will just reduce to metric $f(R)$ gravity and, as mentioned before, it will be equivalent to a Brans--Dicke theory
with $\omega_0=0$.

We have summed up the results presented in this section in a schematic diagram: \\\\
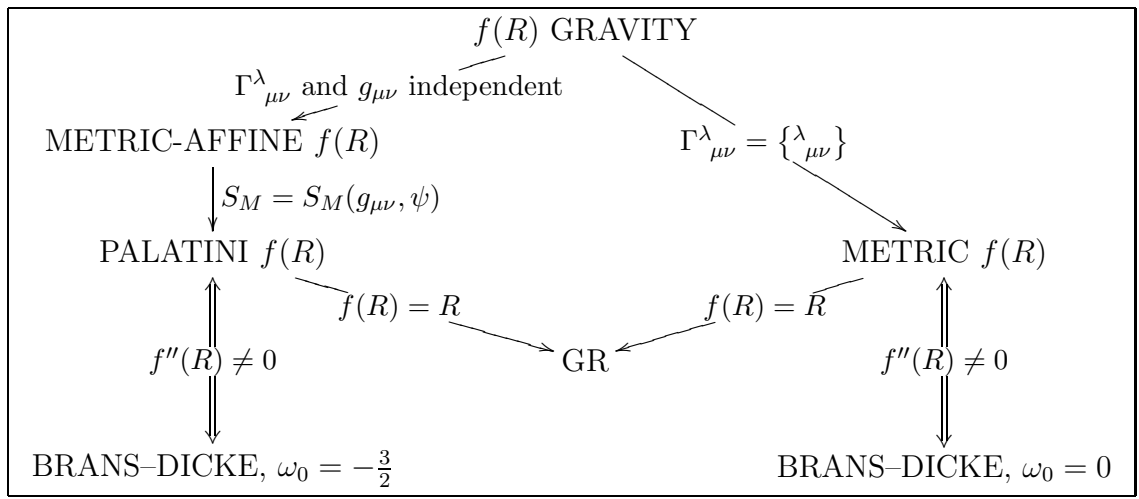
\begin{tabular}{|c|}\hline  $\xymatrix{& \textrm{$f(R)$ GRAVITY} \ar[dl]|{\textrm{{\small $\Gamma^{\lambda}_{\phantom{a}\mu\nu}$ and $g_{\mu\nu}$ independent}}}\ar[ddr]|{\textrm{{\small $\Gamma^{\lambda}_{\phantom{a}\mu\nu}=
%\frac{1}{2}g^{\lambda\sigma}(\partial_\nu g_{\mu\sigma}+\partial_\mu g_{\nu\sigma}-\partial_{\sigma} g_{\mu\nu})
\left\{^\lambda_{\phantom{a}\mu\nu}\right\}
$}}} & \\ \textrm{METRIC-AFFINE $f(R)$} \ar[d]^{{\small \textrm{$S_M=S_M(g_{\mu\nu},\psi)$}}}& &  \\
%\textrm{metric-ar[rrr]^{\textrm{$S_M=S_M(g_{\mu\nu},\psi)$}}
 \textrm{PALATINI $f(R)$}\ar@2{<->}[dd]|{\textrm{{\small $f''(R)\neq 0$}}}\ar[dr]|{\textrm{{\small $f(R)=R$}}}&  & \textrm{METRIC $f(R)$} \ar@2{<->}[dd]|{\textrm{{\small $f''(R)\neq 0$}}}\ar[dl]|{\textrm{{\small $f(R)=R$}}} \\
&  \textrm{GR} &\\
\textrm{BRANS--DICKE, $\omega_0=-\frac{3}{2}$}& &\textrm{BRANS--DICKE, $\omega_0=0$}} 
$\\\hline\end{tabular}

\section{Physical interpretation and implications}
\label{inter}

The fact that, $f(R)$ gravity in the Palatini formalism is equivalent to a class of Brans--Dicke theories
when the matter action is independent of the connection, demonstrates clearly that the former is
intrinsically a metric theory. This could have been expected since the matter is coupled only to
the metric. Even though $\Gamma^{\lambda}_{\phantom{a}\mu\nu}$ is not a scalar, the theory actually
has only one extra scalar degree of freedom with respect to General Relativity \footnote{cf. \cite{tomi} where similar conclusions on the role of the independent connection in $f(R)$ gravity in the Palatini formalism are derived by examining energy conservation}. 
The Jordan frame just prevents us 
from seeing that directly, because the action is written in this frame in terms of
 what turns out to be
an unfortunate choice of variables. On the contrary, if one wants to construct a metric-affine theory of gravity 
matter should be coupled to the connection, as also claimed in \cite{sotlib}. In this case, any dynamical
equivalence with Brans--Dicke theory breaks down. For this reason, it is preferable to reserve the term
metric--affine $f(R)$ theories of gravity for these theories, in order to distinguish them from those for which there is no coupling between the matter 
and the connection and which are usually referred to in the literature as $f(R)$ theories of gravity in the Palatini formalism.

It is also important to mention that $f(R)$ theories of gravity in the metric formalism and in the Palatini formalism
are dynamically equivalent to different classes of Brans--Dicke theories. This implies that they cannot 
be dynamically equivalent to each other. Therefore, these theories will give distinct physical predictions. 
The same is, of course, true for metric-affine $f(R)$ theories of gravity as well, since they cannot be cast into the form of a Brans--Dicke theory. There is, however,
an
exception: metric--affine $f(R)$ gravity will reduce to $f(R)$ theories of gravity in the Palatini formalism
in vacuum, or in any other case where, the only matter fields present are by definition independent
of the connection, 
such as scalar fields, the electromagnetic field or a perfect fluid \cite{sotlib}. Therefore, even though there is no equivalence
between metric-affine $f(R)$ gravity and Brans--Dicke theories with $\omega_0=-3/2$, their phenomenology will be identical 
in many interesting cases, such as cosmological applications.

It should be mentioned that Brans--Dicke theory with $\omega_0=-3/2$ has not 
received very much attention (see however \cite{fabris}). 
The reason for that is that when Brans--Dicke theory was first introduced,
only the kinetic term of the scalar field was present in the action. Therefore, choosing $\omega_0=-3/2$ 
would lead to an ill-posed theory, since only matter described by a stress-energy tensor with a vanishing trace 
could be coupled to the theory. This can be understood by examining eq.~(\ref{bdf3}) in the absence
of terms including the potential. However, once the potential of the scalar field is considered in the action,
no inconsistency occurs. Note that a Brans--Dicke gravitational action with $\omega_0=-3/2$ and no potential term
is conformally invariant and dynamically equivalent with conformal relativity (or Hoyle-Narlikar theory)
 \cite{bla}. The action of conformal relativity \cite{conf1,conf2,conf3} has the form
\be
\label{craction}
S_{CR}=\frac{1}{2\kappa}\int d^4 x \sqrt{-g} \Psi(\frac{1}{6} \Psi R -\Box \Psi).
\ee 
A field redefinition $\Psi^2=6\Phi$ will give
\be
\label{craction2}
S_{CR}=\frac{1}{2\kappa}\int d^4 x \sqrt{-g} \left(\Phi R+\frac{3}{2\Phi}\nabla_\mu \Phi \nabla ^\mu \Phi\right),
\ee 
where a total divergence has been discarded. The dynamical equivalence is therefore straightforward. 

The case of a vanishing potential has no analogue in $f(R)$ gravity. Using eq.~(\ref{defV}) and remembering that $\Phi=f'(\chi)$ and on shell $\tilde{R}=\chi$,
one can easily verify that setting
$V(\Phi)=0$ will lead to the the following equation for $f(\tilde{R})$:
\be
f'(\tilde{R}) \tilde{R}-f(\tilde{R})=0.
\ee
This equation can be identically satisfied only for $f(\tilde{R})=\tilde{R}$. However, to go from the Jordan to the J--H frame 
one assumes $f''(\tilde{R})\neq 0$, so no $f(\tilde{R})$ Lagrangian can lead to a Brans--Dicke theory with a vanishing potential.
It is remarkable that this ill-posed case does not exist in $f(R)$ gravity in the Palatini formalism. 
There is, however, a conformally invariant gravitational action in this context as well. One has to choose
$f(\tilde{R})=a \tilde{R}^2$, where $a$ is some constant \cite{higgs,fer}. 
 In this case the potential of the equivalent Brans--Dicke 
theory will be $V(\Phi)=\Phi^2/4a$. 
For this potential all terms apart from the one containing $T$ in eq.~(\ref{bdf3}) will again vanish, as would
happen for a vanishing potential. The correspondence can easily be generalized 
for $n$-dimensional manifolds, where $n\geq 2$. For the gravitational action to be conformally invariant
in the context of $f(R)$ gravity, one should choose $f(\tilde{R})=a \tilde{R}^{n/2}$ \cite{fer}. 
The corresponding potential can be computed using eq.~(\ref{defV}) and has the form 
\be
V(\Phi)=\left(\frac{n}{2}-1\right)a \left(\frac{2\Phi}{n a}\right)^{n/(n-2)}
\ee 
Eq.~(\ref{bdf2}) will generalize for $n$ dimensions in the following way:
\be
\label{bdf2n}
(n-2)\left(\omega_0+\frac{n-1}{n-2}\right) \Box \phi= \kappa T+\left(\frac{n}{2}-1\right)
\left(\phi V'-\frac{n}{n-2}V\right),
\ee
implying that for $n$ dimensions the special case which we are examining corresponds to $\omega_0=-(n-1)/(n-2)$.
This indicates that a Brans--Dicke gravitational action with $\omega_0=-(n-1)/(n-2)$ and 
a potential $V(\Phi)=b \Phi^{n/(n-2)}$, where $b$ is some constant,
will be conformally invariant in an $n$-dimensional manifold.
As an example we can examine the $4$-dimensional action:
\be
\label{bdactionphi2}
S_{n=4}=\frac{1}{2\kappa}\int d^4 x \sqrt{-g} \left(\Phi R+\frac{3}{2\Phi}\nabla_\mu \Phi \nabla ^\mu \Phi-b \Phi^2\right).
\ee 
Using the redefinition $\Psi^2=6\Phi$ as before, we can bring the action to the following form
\be
\label{cractionpot}
S_{n=4}=\frac{1}{2\kappa}\int d^4 x \sqrt{-g} \Psi(\frac{1}{6} \Psi R -\Box \Psi-b\frac{\Psi^3}{36}),
\ee 
which is a generalization of the action (\ref{craction}). It is easy to verify that this specific potential will not break 
conformal invariance. Under the conformal transformation 
$g_{\mu\nu}\rightarrow \Omega^2 g_{\mu\nu}$ the root of the
determinant will transform as $\sqrt{-g}\rightarrow \Omega^4 \sqrt{-g}$, so with an appropriate redefinition 
of the scalar field, $\tilde{\Psi}=\Omega^{-1}\Psi$ the action will return in the form of (\ref{cractionpot}).

In our study, we have used the Jordan and J--H frames. However, as we said earlier, the results presented so far 
are independent of whether it is the Jordan or the Einstein frame that should be considered as the physical one.
The justification for this claim is related to the mathematical equivalence of frames. Suppose that a theory written in 
the Jordan frame is equivalent to some other theory written also in the same frame; then the mathematical equivalence
between the Jordan and Einstein frames indicates that the Einstein frame versions of these theories will also
 be dynamically equivalent to each other. We also mentioned that the results will remain unchanged if we 
allow the connection present in the action (\ref{palaction}) to be non-symmetric. In the case of
$f(R)$ gravity where the matter is independent of the connection, this is true since the non-symmetric
 part of the connection vanishes even if no such assumption is made {\it a priori}, and the field equations
of the corresponding theory are identical to eqs.~(\ref{field1sym}) and (\ref{field2sym}) \cite{sotlib}. 
On the other hand, when studying the case where matter is coupled to the connection,
we did not have to use the symmetry of the connections, neither did we have to use the field equations, but we
worked at the level of the action.

\section{Equivalence with Brans--Bicke and confrontation with observations}

A very important aspect of the dynamical equivalence of different theories is that it can be used
to apply results derived in the framework of one of the theories to another. This fact was used in 
\cite{chiba} to show that the model of metric $f(R)$ gravity proposed in \cite{carroll0} cannot pass
the solar system tests, since the equivalent Brans--Dicke theory does not. In \cite{cap3} the same prescription
was used to place bounds on the form of the gravitational action of metric $f(R)$ gravity. In $f(R)$
theories of gravity in the Palatini formalism things are more complicated. The exceptional nature of
the equivalent Brans--Dicke theory with $\omega_0=-3/2$ prevents us from using the standard 
Parametrized Post-Newtonian (PPN) expansion \cite{cwill}. The PPN expansion of such theories was computed in \cite{olmo}. 
Confrontation with observations shows that, for a viable theory, $f(R)$ has to be linear to high precision, for 
densities relevant to these tests and therefore any non-linear terms should be suppressed by very small coefficients \cite{olmo,sot1,gianl1}.
 
One cannot follow a similar approach in order to address the same problem for metric-affine gravity, 
 since there is no equivalence with Brans--Dicke theories in this case. One is, however, tempted to use the fact that metric-affine gravity will effectively
reduce to an $f(R)$ theory in the Palatini formalism for specific matter configurations, as mentioned previously. Before continuing the discussion about solar system tests, let us examine this possibility in the case of the Newtonian limit. A theory is said to have a good Newtonian
limit if one can derive the Poisson equation starting from the field equations, in a weak gravity regime and 
when velocities related to matter are negligible compared to the velocity of light. The low velocity assumption is conventionally interpreted as a dominance of the zero-zero component of the stress-energy tensor over all other components. The question one has to answer is how this assumption will affect $\Delta_{\lambda}^{\phantom{a}\mu\nu}$ (eq.~{\ref{field2sym}). In a macroscopic description of matter, a non-vanishing $\Delta_{\lambda}^{\phantom{a}\mu\nu}$ is related to characteristics like heat flows, vorticity, viscosity, etc. \cite{sotlib}. Therefore, we can infer that under the low velocity and weak gravity assumptions, considering a vanishing $\Delta_{\lambda}^{\phantom{a}\mu\nu}$ can be an adequate approximation, at least to first order. In this case, the Newtonian limit of a metric-affine $f(R)$ theory of gravity will coincide with that of the $f(R)$ theory in the Palatini formalism \cite{sot1} with the same choice of the function $f$.

Notice, however, that  the Newtonian limit of a theory discussed above should not be
confused with the solar system tests, as often happens.
Those are experiments performed in the solar system, relevant to
light deflection, radar effects (Shapiro time delay) etc. Therefore, spacetime is taken to be essentially vacuum 
but small corrections related to matter have to be taken into account (for radar experiments, for example, one has to take into account
solar winds) \cite{cwill,grav}. 
They are much more sophisticated tests. 
As a first approximation, 
matter is often treated in these tests as a perfect fluid, since corrections due to anisotropic stresses, for example,
are negligible with respect to other post-Newtonian corrections. Therefore, one might think of using the fact that under
this assumption metric--affine $f(R)$ gravity is effectively equivalent to a Brans--Dicke theory. However, a more rigorous approach is needed in this case. 
When dealing with metric--affine $f(R)$ gravity any deviation from a perfect fluid description of matter, independently of how small it might be,
 will not only lead to small corrections in the PPN expansion but will actually break the equivalence,
falsifying the whole methodology. Therefore, such an approach should be avoided.

 One should, instead, 
form a PPN expansion for metric-affine gravity and then proceed to confrontation
with observations, an analysis which is still pending. It should be noted that this is not an easy task, since it does not constitute a straightforward generalization of the standard PPN expansion. In this case, besides the metric and the stress-energy tensor, there are also other quantities that one would need to expand in post-Newtonian orders, such as the independent connection and $\Delta_{\lambda}^{\phantom{a}\mu\nu}$. Therefore, the most outstanding problem seems to be to find the optimum way to achieve that. One should also mention that the standard PPN formalism is very much simplified by using a freely falling coordinate system, which allows the expansion to be performed around flat spacetime. In metric-affine gravity the independence of the metric and the connections seems to introduce difficulties for defining such a coordinate system. It is interesting to study in which cases the second field equation, eq.~(\ref{field2sym}), can be solved to give the connection in terms of the metric, since in this case one could replace the result in  eq.~(\ref{field1sym}), and end up with only one equation depending on the metric and the matter fields. Then the post-Newtonian expansion could probably be performed in the usual way by expanding the metric and an effective stress-energy complex. Let us close by recalling that metric-affine $f(R)$ theories of gravity reduce in vacuum 
to GR with a cosmological constant which might be a first indication
that numerous models based on them can pass the solar system tests. This allows us to be optimistic, without of course providing a definite answer. It is probable
that when one takes into account corrections relevant to matter, such as the ones mentioned before, this picture might be reversed. 

\section{Conclusions}
\label{concl}

We have studied the relation between scalar--tensor theories of gravity and $f(R)$ theories of gravity in which the
connections are independent of the metric. It has been shown that whether the matter is coupled to 
this independent connection or not, is a crucial aspect of the theory. If there is such a coupling, the theory
is a metric-affine theory of gravity and this connection has its ordinary physical meaning. It is related to parallel
transport and defines the covariant derivative. If there is no coupling between the matter and the independent 
connection, then the theory is equivalent to a Brans--Dicke theory with Brans--Dicke parameter $\omega_0=-3/2$.
In this case, the theory is just a metric theory written using an unfortunate choice of variables which hides
the true degrees of freedom at the level of the action. It is, however, a distinct theory from $f(R)$ theories of gravity in the metric fomalism, which, as already known, are equivalent to 
a Brans--Dicke theory with Brans--Dicke parameter $\omega_0=0$.

It is interesting that even though true metric--affine $f(R)$ gravity cannot be cast in the form of
a Brans--Dicke theory, it effectively reduces to a theory dynamically equivalent to that, when the only matter fields
present are ones whose action is by definition independent of the connection (such as a scalar field, or a perfect fluid).
This shows that the two theories will have the same phenomenology in most applications in cosmology and astrophysics.

Using the equivalence of $f(R)$ gravity and Brans--Dicke theory, we were also able to show that, in $n$-dimensions,
a Brans--Dicke gravitational action with Brans--Dicke parameter $\omega_0=-(n-1)/(n-2)$ and potential $V\propto \phi^{n/(n-2)}$, is conformally invariant.
Finally, we should mention that, since there is no equivalence between Brans--Dicke theory and true metric--affine $f(R)$ gravity,
 one cannot derive a conclusion about whether certain models
will pass the solar system tests, by using a Parametrized Post-Newtonian expansion of Brans--Dicke theories, as done for purely metric
 $f(R)$ gravity, or $f(R)$ gravity in the Palatini formalism. 

\section*{Acknowledgements}

The author is grateful to Stefano Liberati and John Miller for interesting comments and constructive suggestions
during the preparation of this paper.

\end{document}